\documentclass{casc}
\usepackage{algorithm}
\usepackage{algorithmic}
\usepackage{amssymb}
\usepackage{graphicx}

\newcommand{\K}{\mathbb{K}}
\newcommand{\N}{\mathbb{N}}
\newcommand{\M}{\mathbb{M}}
\newcommand{\R}{\mathbb{R}}

\newcommand{\lcm}{\mathop{\mathrm{lcm}}\nolimits}
\newcommand{\lm}{\mathop{\mathrm{lm}}\nolimits}
\newcommand{\lt}{\mathop{\mathrm{lt}}\nolimits}
\newcommand{\pol}{\mathop{\mathrm{pol}}\nolimits}
\newcommand{\anc}{\mathop{\mathrm{anc}}\nolimits}
\newcommand{\nmp}{\mathop{\mathrm{nmp}}\nolimits}

\newcommand{\ba}{\begin{array}{c}}
\newcommand{\ea}{\end{array}}

\def\be{\begin{equation}}
\def\ee{\end{equation}}
\def\ba{\begin{array}{c}}
\def\ea{\end{array}}

\def\ben{$$}
\def\een{$$}

\begin{document}
\title{\bf On Exact Solvability of Anharmonic
Oscillators in Large Dimensions}
\author{Vladimir Gerdt$\,^1$, Denis Yanovich$\,^1$  \and Miloslav Znojil$\,^2$}
\institute{Laboratory of Information Technologies\\
        Joint Institute for Nuclear Research\\
        141980 Dubna, Russia \\
        \email{gerdt@jinr.ru}\ \ \email{yan@jinr.ru} \\[0.3cm]
\and
\'{U}stav jadern\'e fyziky AV \v{C}R\\
250 68 \v{R}e\v{z}, Czech Republic \\
\email{znojil@ujf.cas.cz}}
\date{}
\maketitle
\begin{abstract}
General Schr\"{o}dinger equation is considered with a central
polynomial potential depending on $2q$ arbitrary coupling
constants. Its exceptional solutions of the so called Magyari type
(i.e., exact bound states proportional to a polynomial of degree
$N$) are sought. In any spatial dimension $D \geq 1$, this problem
leads to the Magyari's system of coupled polynomial constraints,
and only purely numerical solutions seem available at a generic
choice of $q$ and $N$. Routinely, we solved the system by the
construction of the Janet bases in a
degree-reverse-lexicographical ordering, followed by their
conversion into the pure lexicographical Gr\"obner bases. For very
large $D$ we discovered that (a) the determination of the
``acceptable" (which means, real) energies becomes extremely
facilitated in this language; (b) the resulting univariate
``secular" polynomial proved to factorize, utterly unexpectedly,
in a {\em fully non-numerical} manner. This means that due to the
use of the Janet bases we found a new exactly solvable class of
models in quantum mechanics.
\end{abstract}

\section{Anharmonic Oscillators and the Problem of Their Solution}

Elementary Hamiltonian $H = p^2 + q^2 + \lambda\,q^4$ of the so
called anharmonic oscillator in one spatial dimension $D = 1$ is an
example which plays a key role in quantum theory and in many of its
applications. We may recollect, for illustration, that small
experimental irregularities in the vibrational spectra in atomic
physics are currently being attributed to the quartic anharmonicity
at a suitable and, if possible, reasonably small coupling constant,
$\lambda = {\cal O}(1)$ \cite{Cizek}. For the fit of some
experimental data of this type one may even employ the two-parametric
family of the Hamiltonians $H = p^2 + q^2 + \lambda\,q^4 +
\varrho\,q^6$ \cite{Singh}, etc. In all these cases, sophisticated
perturbation calculations are usually employed in order to achieve an
agreement between experiment and theory (cf., again, ref.
\cite{Cizek} and many other papers cited therein).

In a mathematically more ambitious setting, Magyari \cite{Magyari}
was probably the first who noticed that in one dimension, Schr\"{o}dinger
equation admits non-perturbative, exceptional but exact bound-state
solutions $\psi^{(Magyari)}(q)$  for {\em any} anharmonic potential
of the following special polynomial form symmetric with
respect to the origin,
 \begin{equation}
 V^{[q]}(r) =
 g_0\,r^2+g_1\,r^4 + \ldots + g_{2{q}}\,r^{4{q}+2}\,,\ \ \ \ \
 \ \ \ \ \ \ \ g_{2q}= \gamma^2 > 0\,,
 \label{geSExt}
 \end{equation}
provided only that its couplings $g_j$ satisfy certain $q$
constraints. These constraints have the form of the system of coupled
polynomial equations (their form will be displayed and discussed
below). Unfortunately, the achievement of the practical compatibility
of the couplings with the Magyari's constraints requires the solution
of his equations by a suitable more or less purely numerical
technique. The corresponding algorithm is usually based on the use of
Gr\"{o}bner bases \cite{Buch85}. The procedure is very standard and
one would have no particular reason for its study in more detail in
general.

The first of the changes which proved relevant in this context
appeared with the introduction of the higher-dimensional
Schr\"{o}dinger equations with polynomial interactions {\em and} with
the Magyari-type solutions \cite{singular}. For all of these models,
the Magyari-type equations become dependent on the dimension $D \geq
1$ playing the role of a new formal parameter. The freedom in its
choice will prove most relevant in our present paper but in a
historical perspective, it still took many years before this chance
has been conceived and described in ref. \cite{quartic} where the
choice of the potential proved restricted, for purely technical
reasons, to the ``first nontrivial" polynomial of type (\ref{geSExt})
with exponent $q=2$.

The main inspiration of our present study of polynomial oscillators with
$q > 2$ lies in the broad and not yet fully explored variety of the
possibilities hidden in a consequent formal analysis of the Magyari-type
equations. In this sense, the next decisive step has been made in refs.
\cite{nje3} where the Magyari's equations proved tractable in semi- and/or
non-numerical manner at the first few lowest choices of the degree of the
wave function, viz., at $N = 1$, $N=2$, and $N=3$. In these ``trivial"
cases, an overall tendency emerged of a distinct separation between the
real (= physical) and complex (= apparently fully redundant) Magyari's
couplings. This result offered an important hint for a more general
analysis of the problem \cite{onitri,G02,GY03} and suggested the idea of
using the Janet bases \cite{Janet,BCGPR1,BCGPR2} in the similar cases. The
simplicity of these low$-N$ models  enabled us to see that an overall and
more systematic study should be directed towards the domain of the very
large $D \gg 1$ (cf. also refs. \cite{Dubna,RS,Gemp,papI} in this
respect).

In spite of the unique success of the {\em mathematics} of
Magyari's nonlinear algebraic equations, a number of
difficulties remained connected with their {\em practical}
applications and applicability at the finite $D$.  One of the
key reasons (and differences from the harmonic oscillator and
other exactly solvable models) is that the {\em explicit
construction} of the Magyari's energies {\em remains purely
numerical}.  Indeed, these values (as well as the related
couplings - we shall show some technical details below) must be
computed as roots of a certain ``secular" polynomial.  This
means that the difference between the variational, ``generic
$N=\infty$" rule in Hilbert space seems only marginally
simplified by the Magyari-type construction of any $N \gg 1$
bound state.

The main purpose of refs. \cite{quartic,RS} derived precisely from
the latter point. Using the idea of perturbation expansions for
Hamiltonians $H=H^{(q,N)}(D)$, these studies proceeded in two steps.
Firstly, a zero-order approximations $H_0^{(q,N)}(\infty)$ have been
constructed while, secondly, a series of corrections has been
evaluated at each particular finite and fixed dimension $D < \infty$.
This opened the market for the constructions of the Hamiltonians
$H_0^{(q,N)}(\infty)$ in systematic manner.

In our present notation, the {exact solvability} of the zero-order
Hamiltonians $H_0^{(q,N)}$ emerged as an utterly unexpected result of
our calculations at $q=1$ in ref. \cite{RS}, at $q=2$ in ref.
\cite{quartic} and at $q=3$ in ref. \cite{papI}.  In what follows we
intend to address the next, more sophisticated problems with $q > 3$.
An emphasis is to be put on the vital role of the methods which were
able to produce the necessary final results {\em within the strict
bound given by the not too fancy available computers}.  Hence, in
what follows, the main emphasis will be laid upon the quality of the
underlying software. Still, a more detailed introductory chapter is
due first.

\section{The Derivation of the Magyari Equations}

\subsection{Harmonic Oscillator with $q=0$ as a Methodical Guide}

The partial differential
Schr\"{o}dinger equation for harmonic oscillator in $D$ dimensions reads
\begin{equation}
\left (
 -
\frac{\hbar^2}{2m} \triangle + \frac{1}{2}\,m\,\Omega^2\,|\vec{x}|^2
\right ) \Psi(\vec{x}) = \varepsilon\, \Psi(\vec{x}) \
 \label{HOgeneric}
\end{equation}
and is solvable by the separation of variables in several systems of
coordinates. The most common cartesian choice may be recommended
for the first few lowest spatial dimensions $D$ only
\cite{Fluegge3}. In contrast, the separation in spherical system
remains equally transparent at any $D$ because it reduces eq.
(\ref{HOgeneric}) to {\em the same} ordinary (so called radial)
differential equation
 \begin{equation}
 \left[-\,\frac{d^2}{dr^2} + \frac{\ell(\ell+1)}{r^2} +
\omega^2r^2
 \right]\, \psi(r) =
E\, \psi(r)
 \label{rad}\
 \end{equation}
with  $r = |\vec{x}| \in (0,\infty)$,  $E
={2m\varepsilon}/{\hbar^2}$ and $ \omega={m\Omega}/{\hbar}>0$. In
this language we have $\ell = \ell_L= L + {(D-3)}/{2}$ where $L =
0, 1, \ldots $. At each $L$ the energy levels are
numbered by the second integer,
 \begin{equation}
E = E_{n,L}=\omega\,(2n+\ell_L+3/2), \ \ \ \ \ \
 \ \ \ \ \ \
 n,L = 0,
1, \ldots\  \label{closeden}.
 \end{equation}
The wave functions with quadratic $ \lambda(r)=\omega\,r^2/2>0$ and
minimal $N = n+1$ in
 \begin{equation}
 \psi_{n,L}(r) =r^{\ell+1}\,
 e^{-\lambda(r)}\,\sum_{m=0}^{N-1}h_m\,r^{2m} \
  \label{ansatz}
 \end{equation}
are proportional to an $n$th Laguerre polynomial \cite{FluLag}. In Hilbert
space, their set is complete.

\subsection{ $q=1$ and Quasi-Exact (i.e., Incompletely Solvable)
 Sextic Oscillators}

An immediate partially or quasi-exactly solvable (QES)
generalization of harmonic oscillators was discovered by Singh et al
\cite{Singh}. In this case one replaces
 \begin{equation}
 \omega \longrightarrow W(r)
  = \alpha_0 +\alpha_1r^2 \,, \ \ \ \ \ \ \
 V^{(HO)}(0) \longrightarrow  G_{-1}+ G_0r^2 = U(r)\,
 \label{omega}
 \end{equation}
and gets the general sextic potential
 \begin{equation}
 V^{(sextic)}(r) =U(r)+ r^2W^2(r)=
 g_0\,r^2+g_1\,r^4 + g_{2}\,r^{6}\,
 \label{SExta}
 \end{equation}
all three couplings of which are simple functions of our initial
three parameters and {\it vice versa},
 \begin{equation}
 \ \ \ g_{2}= \alpha_1^2 > 0,
 \ \ \ g_{1}= 2\,\alpha_0\alpha_1,
 \ \ \  g_{0}= 2\,\alpha_0^2+G_0
 , \ \ \ G_{-1}=0\,.
 \label{mapa}
 \end{equation}
The resulting Schr\"{o}dinger bound state problem cannot be solved
in closed form. Nevertheless, we may postulate the polynomiality
of the wave functions $ \psi^{(sextic)}_{n,L}(r)$ {\em for a
finite multiplet} (i.e., $N-$plet) of the wave functions. Under
 the specific constraint
 \begin{equation}
 G_0=-\alpha_0^2-\alpha_1(4N+2\ell+1), \ \ \ \ \ \ N \geq 1
 \label{lowers}\
  \end{equation}
this $N-$plet of polynomial solutions (\ref{ansatz}) is
made exact by the choice
of a WKB-like (i.e., quartic) exponent
 \begin{equation}
  \lambda(r) =
 \frac{1}{2}{\alpha}_0 r^2 + \frac{1}{4}{\alpha}_1 r^4\,.
 \label{qrawS}
 \end{equation}
The ansatz (\ref{ansatz})
transforms then the differential Schr\"{o}dinger equation into a
linear algebraic definition of the unknown $N-$plet of coefficients
$h_m$. The solution is always obtained for a mere finite
set of the levels $n \in (n_0, n_1, \ldots,n_{N-1})$. In contrast to
the harmonic oscillator, the QES solvability is based on the $L-$ and
$N-$ dependent constraint (\ref{lowers}) so that, generically, the
elementary QES multiplet exists in a single partial wave only.

\subsection{Magyari's QES Oscillators with $q > 1$}

The explicit energy formula (\ref{closeden}) for harmonic oscillator
was replaced by an implicit definition in the preceding paragraph
which gives the sextic QES energies {\em in the purely numerical
form}, viz., as zeros of the Singh's secular determinant of a certain
tridiagonal $N$ by $N$ matrix \cite{Ushveridze}.
In this sense,  Magyari \cite{Magyari}
generalized
 the Singh's QES construction.
In our present notation we may put, simply,
 $$
 V^{(q)}(r) =U^{(q)}(r)+ r^2[W^{(q)}(r)]^2\,,
 \ \ \ \ \ \ \  U^{(q)}(r) = G_0r^2 +  G_1r^4 +\ldots + G_{q-1}r^{2q}
 \,,
  $$
  \begin{equation}
 \ \ \ \ \ \ \
W^{(q)}(r)
  = \alpha_0 +\alpha_1r^2 + \ldots + \alpha_q\,r^{2q}\,
 \label{greeSEx}
 \end{equation}
This formula re-parametrizes the polynomial (\ref{geSExt})
and specifies the one-to-one correspondence between the two sets
of couplings,
 $$
\{g_0, \ldots, g_{2q} \} \Longleftrightarrow \{G_0, \ldots,
G_{q-1}, \alpha_0, \ldots, \alpha_{q} \}\,
 $$
where $g_{2q}= {{\alpha}_{q}}^2$, $g_{2q-1} =g_{2q-1}
(\alpha_{q},\alpha_{q-1})= 2\,{\alpha}_{q-1}\,{\alpha}_{q}, \ldots$
or, in opposite direction, ${\alpha}_{q}=\sqrt{g_{2q}}\equiv \gamma
>0$, ${\alpha}_{q-1}=g_{2q-1}/{(2\alpha_q)}$ etc.

At any $q=1, 2, \ldots$, equation (\ref{qrawS}) must be further
modified,
 \begin{equation}
 \lambda^{(q)}(r) =
 \frac{1}{2}{\alpha}_0 r^2 + \frac{1}{4}{\alpha}_1 r^4 +
\ldots + \frac{1}{2q+2} {\alpha}_{q} r^{2q+2}
 \label{qraw} \label{Jost}\,.
 \end{equation}
With $\alpha_q > 0$, one verifies that
 $$
 \psi^{(physical)}(r) \approx e^{-\lambda^{(q)}(r)+{\cal O}(1)},\
 \ \ \ \   r \gg 1\,
 $$
which means that the correct bound-state ansatz
 \begin{equation}
\psi(r) = \sum_{n=0}^{N-1}\, h_n^{(N)} \,r^{2n+\ell+1}\,{\rm
exp}\left [ -\lambda^{(q)}(r)\right ]
 \label{ana}
 \end{equation}
converts our radial equation (\ref{rad}) + (\ref{geSExt}) into an
equivalent linear algebraic problem
\begin{equation}
\hat{Q}^{[N]}\,\vec{h}^{(N)}=0
 \label{tridSE}
 \end{equation}
with an asymmetric and {\em non-square} matrix
 \begin{equation}  \hat{Q}^{[N]}=
 \left(
  \begin{array}{lllllll}
 B_{0} & C_0&  & & && \\
 A_1^{(1)}&B_{1} & C_1&    &&& \\
 \vdots&&\ddots&\ddots&&&\\
 A_q^{(q)}& \ldots& A_q^{(1)} &
 B_{q} & C_q&    & \\
 &\ddots&&&\ddots&\ddots&\\
 &&A_{N-2}^{(q)}& \ldots & A_{N-2}^{(1)}&
 B_{N-2} & C_{N-2} \\
 &&&A_{N-1}^{(q)}& \ldots & A_{N-1}^{(1)}&
 B_{N-1} \\
 &&&&\ddots&\vdots&\vdots\\
 &&&&&A_{N+q-2}^{(q)}&A_{N+q-2}^{(q-1)}\\
 &
 &&&&&A_{N+q-1}^{(q)}\\
 \end{array}
 \right ).
 \end{equation}
Its elements depend on the parameters in bilinear manner,
\begin{equation}
\begin{array}{c}
   C_n = (2n+2)\,(2n+2\ell+3),\ \ \ \ \ \
  B_n = E-\alpha_0\,(4n+2\ell +{3})
\\
 A_n^{(1)} = -\alpha_1\,(4n+2\ell+1) + \alpha_0^2 -g_0, \ \ \ \ \ \ \
 A_n^{(2)} = -\alpha_2\,(4n+2\ell-1) + 2 \alpha_0\alpha_1 -g_1,  \\
 \ldots,\\
 A_n^{(q)} = -\alpha_q\,(4n+2\ell+3-2q) +
 \left (\alpha_0\alpha_{q-1}+ \alpha_1\alpha_{q-2}+ \ldots
  +\alpha_{q-1}\alpha_0
 \right ) -g_{q-1},
 \\
\ \ \ \ \ \ \ \ \ \ \ \ \ \ \ \ \
  \ \ \ \ n = 0, 1, \ldots \ .
  \end{array}
  \label{elem2}
   \end{equation}
At {\em any} fixed and finite $N =1, 2, \ldots$ the non-square system
(\ref{tridSE}) is an over-determined set of $N+q$ linear equations
for the $N$ non-vanishing components of the vector $\vec{h}^{(N)}$.
At $q=0$ these equations degenerate back to the recurrences and
define the harmonic oscillator states. At $q=1$ we return to the
sextic model where the ''redundant" last row fixes one of the
couplings and where we are left with a diagonalization of an $N$ by
$N$ matrix which defines the $N-$plet of the real QES energies in
principle. The situation is more complicated at $ q > 1$. The
counting of parameters and equations indicates that unless one
broadens the class of potentials, only a very small multiplet of
bound states may remain available in closed form \cite{Leach}.

Using an elementary change of variables, one may transform the
decadic forces into their quartic equivalents etc. Paper
\cite{classif} may be consulted for details which indicate that the
study of any potential $V(r)$ {\em which is a polynomial in any
rational power of the coordinate $r$} may be replaced by the study of
its present Magyari's or ''canonical" QES representation $V^{(q)}(r)$
at a suitable integer $q$. In addition, we shall also restrict our
attention to the domain of large $D$.

\section{Magyari Equations
at the Large Spatial Dimensions}

Up to now, our attention has been concentrated upon the structure of
the QES wave functions. From the point of view of the evaluation of
the energies, the main dividing line between the solvable and
unsolvable spectra is in fact marked by the distinction between the
closed $q=0$ formulae and their implicit QES form at $q=1$. The
transition to the next $q=2$ may be perceived as merely technical. At
all $q \geq 1$, the difficulties grow with $N$. In such a setting
the emergence of certain simplifications at $D \gg 1$ may be crucial.

\subsection{An Exceptional, Decoupled Last-Row Constraint}

At any $D$, the last row in eq. (\ref{tridSE}) decouples from the
rest of the system. At any $q>1$ it may treated as a constraint
which generalizes eq. (\ref{lowers}),
 \begin{equation}
 g_{q-1}  = -\alpha_q\,(4n+2\ell+3-2q) +
 \left (\alpha_0\alpha_{q-1}+ \alpha_1\alpha_{q-2}+ \ldots
  +\alpha_{q-1}\alpha_0
 \right ).
 \label{redef}
 \end{equation}
The insertion of this explicit definition of the coupling
$g_{q-1}$ simplifies the lowest diagonal in $\hat{Q}^{[N]}$,
 \begin{equation}
 A_n^{(q)} = 4\,\gamma\,(N+q-n-1).
 \end{equation}
Since $ A_{N+q-1}^{(q)}=0$ we may drop the ''hat" $\hat{\,}$ and
re-write eq. (\ref{tridSE}) in the more compact form where the size
of the non-square matrix $Q^{[N]}$ is merely $(N+q-1)$ by $N$,
\begin{equation}
Q^{[N]}\,\vec{h}^{(N)}=0\ .
 \label{ridiSE}
 \end{equation}
This is the proper Magyari's system and it is merely solvable
non-numerically in the simplest case with $q=0$. No coupling is then fixed
and the energies themselves are given by the explicit formula
(\ref{redef}). Also the recurrences for coefficients of the wave functions
may be solved in compact form.

The next, $q=1$ version of eq. (\ref{ridiSE})
degenerates to the single, determinantal secular equation
 \begin{equation}
 \det Q^{[N]}=0.
 \label{secular}
 \end{equation}
Its solution is a purely numerical problem at all the larger $N \geq 5$.
Of course, one coupling is fixed by eq. (\ref{redef}) and only the
$N-$plet of energies must be calculated as represented by the real zeros
of the single secular polynomial.

At the larger exponents
$q\geq 2$, some $q$ mutually coupled $N$ by $N$
secular determinants must vanish simultaneously \cite{Dubna}. With an
auxiliary abbreviation for the energy $E = -g_{-1}$ this means that
at least one of the couplings is always energy-dependent and that its
value must be determined numerically. In the other words, our
non-square matrix $Q^{[N]}=Q^{[N]}(g_{-1}, g_0, \ldots,g_{q-2})$ will
annihilate the vector $\vec{h}^{(N)}$ if and only if all its $q$
arguments are determined in a deeply nonlinear and self-consistent,
mostly purely numerical manner.

\subsection{Coupled Constraints at $D \gg 1$ \label{reasons}}

In our approach the {\em guaranteed} polynomiality of the wave
functions will play a key role.
One can say that
in our original differential eq. (\ref{rad}) the numerical value
of the spatial dimension $D$ will be assumed large.
No other
simplifications will be assumed.

In our problem with
the old matrix elements
\begin{equation}
\begin{array}{c}
   C_n = (2n+2)\,(2n+2L+D),\ \ \ \ \ \
  B_n = -g_{-1}-\alpha_0\,(4n+2L+D),
\\
 A_n^{(k)} = -g_{k-1} -\alpha_k\,(4n+2L+D-2k) +
 \left (\alpha_0\alpha_{k-1}+  \ldots
  +\alpha_{k-1}\alpha_0
 \right ),\\
\ \ \ \ \ \ \ \
 k = 1, 2, \ldots, q-1,\
 \ \
  \ \ \ \ \
  \ \ \ \ n = 0, 1, \ldots , N+q-2\
  \end{array}
  \label{elemih2}
   \end{equation}
we shall preserve the dominant components
of the matrix elements only,
  $$
     C_n^{[0]} = (2n+2)\,D,\ \ \ \ \ \
  B_n^{[0]} = -g_{-1} -\alpha_0\,D,
\ \ \ \ \ \ \
 A_n^{(k)[0]} = -g_{k-1} -\alpha_k\,D \ ,k < q
  \label{emih2}
   $$
(note that $ A_n^{(q)} = A_n^{(q)[0]}$ is unchanged).
Then we re-scale the coordinates and, hence, coefficients
according to the rule
 \begin{equation}
 h^{(N)}_n=p_n/\mu^n\,.
 \label{muna}
 \end{equation}
Simultaneously we have to replace the energies and couplings
$\{g_{-1}, g_0, \ldots,g_{q-2}\}$ by the new re-scaled parameters $\{
s_1, s_2,\ldots, s_q\}$ using the following linear recipe,
 \begin{equation} \label{recipe}
 g_{k-2}=
 -\alpha_{k-1}D-\frac{\tau}{\mu^{k-1}}\,s_k,
 \ \ \ \ \ \ \
 k = 1, 2, \ldots, q\ .
 \end{equation}
where we abbreviated
 \begin{equation}
 \mu=\mu(D) = \left ( \frac{D}{2\gamma} \right )^{1/(q+1)},
 \ \ \ \ \ \
 \tau=\tau(D) = \left ( 2^{q+2}\,D^{q}\,\gamma
 \right )^{1/(q+1)}
 .\label{mu}
 \end{equation}
In the leading-order approximation this gives, finally,
our Magyari
equations in the compact form
 \begin{equation}
 \label{trap}
 \left( \begin{array}{cccccc}
 s_1 & 1 & & & & \\
 s_2 & s_1 & 2 & &  & \\
 \vdots&  &\ddots & \ddots & & \\
 s_q & \vdots &  &s_1 & N-2& \\
 N-1& s_q & &  & s_{1} & N-1 \\
 &N-2& s_q & & \vdots& s_{1}  \\
 &&\ddots&\ddots&&\vdots\\
 &&& 2 & s_q&  s_{q-1}   \\ &&&& 1 & s_q
\end{array} \right)
 \left( \begin{array}{c}
 {p}_0\\
 {p}_1\\
\vdots \\
 {p}_{N-2}\\
 {p}_{N-1}
\end{array} \right)
=
0\
 \end{equation}
which is to be studied in what follows.

\section{The Method of Solution of the $D \gg 1$ Magyari Equations}

\subsection{Involutive Bases}

To solve polynomial systems (\ref{trap}) we shall construct for them
the related Janet bases.  Janet bases are typical representatives of
general involutive bases of polynomial ideals~\cite{GB1} which are
Gr\"obnerian though, generally, redundant. However, just this
redundancy of involutive bases makes the structural and combinatorial
information on polynomial and differential ideals and modules more
accessible~\cite{Apel98,G99,Seiler1,Seiler2}.

And as well as the reduced Gr\"obner bases, the involutive bases
can be used for solving polynomial systems with finitely
many solutions that correspond to the zero-dimensional
ideals~\cite{CLO}.  For this purpose, a pure lexicographical
monomial order seems best since it provides the completely
triangular basis with sequentially eliminated variables starting
from the highest one with respect to the order
chosen~\cite{Buch85}. However, computation of a lexicographical
basis takes usually much more time than computation of a
degree-reverse-lexicographical basis, first, and conversion of
this basis into the lexicographical one, second.

We use this two-step computational procedure in our study and
solving~(\ref{trap}). In doing so we shall deal with the minimal Janet
bases~\cite{GB2} only. It is remarkable that a
degree-reverse-lexicographical order is inherent in minimal Janet bases of
zero-dimensional ideals. What follows from the demonstration
in~\cite{Seiler1} is this inherence to Pommaret bases, and the fact proven
in~\cite{G00} that a minimal Janet basis is also a Pommaret basis whenever
the latter exists, i.e., whenever it is finite. Zero-dimensional ideals
always have finite Pommaret bases~\cite{Pommaret78}.

\subsection{Janet Bases}

Below we use the definitions and notations from~\cite{GB2,G00,GBY1,GBY2}:
$\N$ is the set of non-negative integers;
$\M=\{x_1^{d_1}\cdots x_n^{d_n} \mid d_i\in \N\}$ is the set of
monomials in the
polynomial ring $\R=\K[x_1,\ldots,x_n]$ over zero characteristic field $\K$;
$\deg_i(u)$ is the degree of $x_i$ in
$u\in \M$; $\deg(u)=\sum_{i=1}^m \deg_i(u)$ is the total degree of
$u$; $\succ$ is an admissible~\cite{Buch85,CLO} monomial
ordering compatible with
$$
x_1\succ x_2\succ\cdots\succ x_n\,.
$$
Divisibility of monomial $v$ by monomial $u$ will be written as
$u\mid v$. A divisor $u$ of a monomial $v$ is proper if
$deg(u)<deg(v)$. $\lm(f)$ and $lt(f)$ denote, respectively, the
leading monomial and the leading term of the polynomial $f\in
\R$ with respect to $\succ$. $\lm(F)$ denotes the leading
monomial set for $F$, and $Id(F)$ denotes the ideal in $R$
generated by $F$.

Let polynomial set $F\subset \R$ be finite and $f\in F$.  For
each $1\leq i\leq n$ we partition $F$ into groups labeled by
non-negative integers
$d_1,\ldots,d_i$:
$$
[d_1,\ldots,d_i]=\{\ f\ \in F\ |\ d_j=\deg_j(\lm(f)),\ 1\leq j\leq i\ \}.
$$
A variable $x_i$ is {\em (Janet) multiplicative}
for $f\in F$ if $i=1$ and
$$\deg_1(\lm(f))=\max\{\deg_1(\lm(g))\ |\ g\in F\},$$
or if
$i>1$, $f\in [d_1,\ldots,d_{i-1}]$ and
$$\deg_i(\lm(f))=\max\{\deg_i(\lm(g))\ |\ g\in
[d_1,\ldots,d_{i-1}]\}.$$
If a variable is not multiplicative
for $f\in F$, it is {\em nonmultiplicative} for $f$ and we write this as
$x_i\in NM_J(f,F)$. $u\in \lm(F)$ is a {\em Janet divisor}
of $w\in \M$, if $u\mid w$ and monomial $w/u$ contains only
multiplicative variables for $u$. In this case we write $u\mid_J
w$.

A finite polynomial set $F$ is {\em Janet autoreduced} if each
term in every $f\in F$ has no Janet divisors among
$\lm(F)\setminus \lm(f)$. A polynomial $h\in \R$ is in the {\em
Janet normal form modulo $F$} if every term in $h$ has no $J-$
divisors in $\lm(F)$. We denote the Janet normal form of
polynomial $f$ modulo $F$ by $NF_J(f,F)$. If the leading
monomial $\lm(f)$ of $f$ has no Janet divisors among elements in
$\lm(F)$, then we say that $f$ is in the {\em Janet head normal
form} modulo $F$ and write $f=HNF_J(f,F)$.

A Janet autoreduced set $F$ is a {\em Janet basis} of $Id(F)$ if
any {\em nonmultiplicative prolongation} ( multiplication by a
nonmultiplicative variable ) of any polynomial in $F$ has
vanishing Janet normal form modulo $F$:
\begin{equation}
 (\forall f\in F)\ (\forall x\in NM_J(f,F))\ \ [\ NF_J(f\cdot
x,F)=0\ ]\,. \label{J_basis}
\end{equation}
A Janet basis $G$ of ideal $Id(G)$ is {\em minimal} if for any
other Janet basis $F$ of the ideal the inclusion $lm(G)\subseteq
\lm(F)$ holds. A monic minimal Janet basis is uniquely defined
by an ideal and a monomial order.  In what follows we deal with
the minimal Janet bases only and often omit the word ``minimal''.

\subsection{Algorithm for Computing Janet Bases}

We present now the algorithm {\bf JanetBasis} which is a special form of
the general Gerdt--Blinkov algorithm~\cite{G02,GB2} for computing minimal
involutive bases concretized for Janet division. This concretization in
its more detailed form relied on the appropriate data structures -- Janet
trees -- and is described in~\cite{GBY1,GBY2}. Note that, recently, the
Gerdt--Blinkov algorithm in its form presented in ~\cite{G02,G99} was
implemented in Maple for both the polynomial and the linear differential
ideals~\cite{BCGPR1,BCGPR2}.

To provide minimality of the output Janet basis~\cite{GB2} the
intermediate data, i.e. initial polynomials and their prolongations and
reductions, are partitioned into two subsets $T$ and $Q$. Set $T$ contains
a part of the intermediate basis. Another part of the intermediate data
contained in set $Q$ also includes all the nonmultiplicative prolongations
of polynomials in $T$ which must be examined in accordance with the
definition of Janet bases.

To apply the involutive
analogues of the Buchberger criteria and to avoid
repeated prolongations we endow with every polynomial
$f\in F$ the triple structure $$p=\{f,\, u,\, vars\}$$
such that
$$
\begin{array}{lcl}
\pol(p)&=&f\ \mbox{is polynomial}\ f\ \mbox{itself},\\
\anc(p)&=&u\ \mbox{is the leading monomial of} ~\mbox{a
polynomial ancestor of}\ f\ \mbox{in}\ F,\\
\nmp(p)&=&vars \ \mbox{is a (possible empty) subset of variables}.
\end{array}
$$
Here the {\em ancestor} of $f$ is a polynomial $g\in F$ with
$u=\lm(g)$ and such that $u\mid \lm(p)$. Moreover, if
$\deg(u)<\deg(\lm(p))$, then every variable occurring in the
monomial $\lm(p)/u$ is nonmultiplicative for $g$. Besides, for
the ancestor $g$ the equality $\anc(g)=\lm(g)$ must hold.  These
conditions mean that polynomial $p$ was obtained from $g$, in
the course of the below algorithm {\bf JanetBasis},  by a
sequence of nonmultiplicative prolongations.  This tracking of
the history in the algorithm allows one to use the involutive
analogues of Buchberger's criteria to detect and avoid
unnecessary reductions.

The set $vars$ contains those nonmultiplicative variables which
have been already used in the algorithm for construction of
nonmultiplicative prolongations.
This set serves to prevent the repeated prolongations.

After every insertion of a new element $p$ in $T$ all elements
$r\in T$ such that $\lm(r)\succ \lm(p)$ are moved from $T$ to
$Q$ in line 13. Such a displacement provides minimality of the
output basis.

It should also be noted that for any triple $p\in T$ the set $vars$
must always be a subset of the set of nonmultiplicative variables of
$\pol(p)$. Line 21 controls this condition.

\begin{algorithm}{\bf Algorithm JanetBasis($F,\prec$)\label{Algorithm}}
\begin{algorithmic}[1]
\INPUT $F\in \R\setminus \{0\}$, a finite polynomial set\\
 \hspace*{1.0cm} $\prec$, an admissible ordering \\
\OUTPUT $G$, a minimal Janet basis of $Id(F)$
\STATE {\bf choose} $f\in F$ with the lowest $\lm(f)$ w.r.t. $\succ$
\STATE $T:=\{f,\lm(f),\emptyset\}$
\STATE $Q:=\{\{q,\lm(q),\emptyset\} \mid q\in F\setminus \{f\}\}$
\STATE $Q:=${\bf JanetHeadReduce}$(Q,T)$
  \WHILE{$Q \neq \emptyset$}
    \STATE {\bf choose} $p \in Q$ such that $\lm(\pol(p))$ has no
         proper divisors
         among $\{\lm(\pol(q)) \mid q\in Q\setminus \{p\} \}$
    \IF{$\lm(\pol(p)) = 1$}
      \RETURN \{1\}
    \ELSE
    \STATE $Q:=Q\setminus \{p\}$
    \IF{$\lm(\pol(p)) = \anc(p)$}
          \FORALL{$\{r \in T \mid lm(\pol(r))\succ \lm(\pol(p))\}$}
          \STATE $Q:=Q \cup \{r\}$; \hspace*{0.4cm} $T:=T \setminus \{r\}$
          \ENDFOR
    \ENDIF
    \STATE $\pol(p):=\mathbf{NF_J}(\pol(p),T)$
    \ENDIF
    \STATE $T:=T \cup \{p\}$
    \FORALL{$q\in T$ {\bf and} $x\in NM_J(\pol(q),T)\setminus \nmp(q)$}
      \STATE $Q:=Q \cup \{\{\pol(q)\cdot x,\anc(q),\emptyset\}\}$
      \STATE $\nmp(q):=\nmp(q)\cap NM_J(\pol(q),T)\cup \{x\}$
    \ENDFOR
    \STATE $Q:=${\bf JanetHeadReduce}$(Q,T)$
  \ENDWHILE
  \RETURN $G:=\{\pol(f)\mid f\in T\}$
\end{algorithmic}
\end{algorithm}

The initialization step is done in lines 1--4. The subalgorithm {\bf
JanetHeadReduce} performs Janet reduction of the leading terms of
polynomials in $Q$ modulo polynomials in $T$.

In the main loop 5--24 an element in $Q$ is selected in line 6. The
correctness of this selection strategy proved in~\cite{G01}. In practice
the cardinality $Q$ at intermediate steps of the algorithm is rather large
and easily runs up to hundreds and thousands. At the same time there may
be different polynomials in $Q$ with identical leading monomials.
Therefore, the restriction in line 6 still admits some arbitrariness. In
our implementation in~\cite{GBY2} for the degree-reverse-lexicographical
ordering a triple $p\in Q$ with the minimal $\deg(\lm(\pol(p)))$ was
chosen. In the case of several such polynomials in $Q$, the one with the
minimal number of terms was picked up.

Line 8 breaks computations in the case when inconsistency is
revealed during the head term reduction in $Q$ and returns the
unit basis. In line 16 the tail Janet reduction is done, then
the Janet reduced polynomial in $p$ is inserted in $T$ and all
the higher ranked polynomials are moved to $Q$ (loop 12-14).
Actually this displacement takes place only if a polynomial in
$p$ has been subjected by the head term reduction in line 23.
Otherwise, $\pol(p)\succ \pol(r)$ holds for any $r\in T$. The
insertion of a new polynomial in $T$ may generate new
nonmultiplicative prolongations of elements in $T$ which are
added to $Q$ in line 20.  To avoid repeated prolongations the
set $\nmp(q)$ of Janet nonmultiplicative variables for $q$ has
been used to construct its prolongations is enlarged with $x$ in
line 21.

The subalgorithm {\bf JanetHeadReduce} computes the Janet head normal form
of polynomials in $Q$ modulo polynomials in $T$

\begin{algorithm}{\bf Subalgorithm JanetHeadReduce($Q, T$)}
\begin{algorithmic}[1]
\INPUT $Q$ and $T$, sets of triples
\OUTPUT Janet head reduced set $Q$ modulo $T$
 \STATE $S:=Q$
 \STATE $Q:=\emptyset$
 \WHILE {$S \neq \emptyset$}
   \STATE {\bf choose} $p\in S$
   \STATE $S:=S\setminus \{p\}$
   \STATE $h:=\mathbf{HNF_J}(p,T)$
   \IF{$h\neq 0$}
       \IF{$\lm(\pol(p))\neq \lm(h)$}
       \STATE $Q:=Q\cup \{h,\lm(h),\emptyset\}$
       \ELSE
       \STATE $Q:=Q\cup \{p\}$
       \ENDIF
   \ENDIF
\ENDWHILE \RETURN $Q$
\end{algorithmic}
\end{algorithm}

\noindent and invokes in line 6 subalgorithm $\mathbf{HNF_J}(p,T)$ that
does head reduction of a single polynomial $p$.

\begin{algorithm}{\bf Subalgorithm $\mathbf{HNF_J}(f,T)$}
\begin{algorithmic}[1]
\INPUT $f=\{\pol(f),\anc(f),\nmp(f)\}$, a triple\\
 \hspace*{1.0cm} $T$, a set of triples
\OUTPUT $h=HNF_J(\pol(f),T)$, the Janet head normal form
   of the polynomial in $f$ modulo polynomial set in $T$
    \STATE $G:=\{\pol(g)\mid g\in T\}$
    \IF{$\lm(\pol(f))$ is involutively irreducible modulo $G$}
       \RETURN $f$
    \ELSE
       \STATE $h:=\pol(f)$
       \STATE {\bf choose} $g\in T$ such that $\lm(\pol(g))\mid_J \lm(h)$
       \IF{$\lm(h) \neq \anc(f)$}
          \IF{{\bf CriterionI}$(f,g)$ {\bf or} {\bf CriterionII}$(f,g)$}
            \RETURN $0$
          \ENDIF
       \ELSE
         \WHILE{$h\neq 0$ {\bf and} $\lm(h)$ is $L-$reducible modulo $G$}
            \STATE {\bf choose} $q\in G$ such that $\lm(q)\mid_J \lm(h)$
            \STATE $h:=h - q\cdot \lt(h)/\lt(q)$
         \ENDWHILE
       \ENDIF
    \ENDIF
  \RETURN $h$
\end{algorithmic}
\end{algorithm}

\noindent
For a head reducible input polynomial $\pol(f)$ the two
involutive analogues of the Buchberger criteria~\cite{Buch85}
criteria are verified in line 8 of subalgorithm
$\mathbf{HNF_J}$:
\begin{itemize}
\item {\bf Criterion I}$(f,g)$ is true iff $\anc(f)\cdot \anc(g)
\mid \lm(\pol(f))$.
\item {\bf Criterion II}$(f,g)$ is true iff
$\deg(\lcm(\anc(f)\cdot \anc(g)))< \deg(\lm(\pol(f))$.
\end{itemize}
If any of the two criteria is true, then
$HNF(\pol(f),T)=0$~\cite{GBY2}. Though as shown
in~\cite{Hemmecke} {\bf Criterion II} does not fully replace the
Buchberger chain criterion, in practice {\bf Criterion II} works
pretty well as our computer experiments demonstrate~\cite{GBY2}.

The last subalgorithm $\mathbf{NF_J}$ performs the Janet tail reduction of a
polynomial with irreducible leading term. It outputs the full Janet
normal form $NF_J(f,T)$ of the input polynomial $f$ modulo polynomial
set containing in $T$. This subalgorithm is called in line 16 of the main
algorithm {\bf JanetBasis} and performs a chain of elementary involutive
reductions until every term in the obtained polynomial becomes Janet
irreducible modulo polynomials in $T$.

\begin{algorithm}{$\mathbf{NF_J}(f,T)$}
\begin{algorithmic}[1]
\INPUT $f$, a polynomial such that $f:=HNF_J(f,T)$; \\
 \hspace*{0.7cm} $T$, a set of triples
\OUTPUT $h=NF_J(f,T)$, the full Janet normal form of $h$ \\
 \hspace*{1.2cm}modulo polynomial set in $T$
    \STATE $G:=\{\pol(g)\mid g\in T\}$
    \STATE $h:=f$
      \WHILE{$h\neq 0$ {\bf and} $h$ has a term $t$ Janet reducible modulo $G$}
            \STATE {\bf choose} $g\in G$ such that $\lm(g)\mid_J t$
            \STATE $h:=h - g\cdot t/\lt(g)$
      \ENDWHILE
  \RETURN $h$
\end{algorithmic}
\end{algorithm}

It should be noted that both the full Janet normal form and
the Janet head normal form are uniquely defined and, hence, uniquely
computed by the above subalgorithms. This uniqueness is a
consequence of a Janet divisor among the leading terms of polynomials
in $T$ at every step of intermediate computations~\cite{GB2}.

\subsection{Converting Bases and Finding Roots}

As we emphasize in the previous section, to find common roots of
polynomials in a given system it is worthwhile to compute a pure
lexicographical involutive or reduced Gr\"obner basis. We do
this computation in the following three steps:

\begin{enumerate}
\item Computation of a minimal degree-reverse-lexicographical
Janet basis by the above described algorithm.
\item Extraction from the Janet basis obtained the reduced Gr\"obner basis.
\item Conversion of the degree-reverse-lexicographical Gr\"obner
basis into the pure lexicographical one by the famous FGLM
algorithm~\cite{FGLM}.
\end{enumerate}

\noindent
Step 2 is done immediately due to the history of prolongations
stored in the polynomial triples (Sect. 4.3).  Since the reduced
Gr\"obner basis is a subset of the Janet basis
computed~\cite{GB2} and this subset is irreducible with respect
to the conventional (noninvolutive) reductions, a triple
$p=\{f,\, u,\, vars\}$ in the Janet basis contains an element
$f$ of the reduced Gr\"obner basis if and only if $lm(f)=u$.
This relation means that an element of the reduced Gr\"obner
basis is such an element in the Janet basis that it has no
ancestors in the last basis. Indeed, the leading term of this
element cannot be a prolongation of the leading term of other
element in the basis. In addition to the use of criteria
(Section 5.2) this is one more byproduct of the triple
representation.

The conversion of the degree-reverse-lexicographical Gr\"obner
basis extracted (old basis) into the pure lexicographical basis
(new basis) is done as follows~\cite{FGLM,CLO2}. First, a
sequence of monomials is generated, starting from the least ones
w.r.t. to the new ordering and then their normal forms are
computed modulo the old basis until there appears a monomial
whose normal form is a linear combination of normal forms of the
preceding monomials. In this case, we add the polynomial given
by this relation to the new basis. This process is continued by
constructing other elements in the new basis by treatment of the
next variables in accordance with the new monomial order.
Computation of the normal form for a monomial is simplified if
one takes into account the fact that the normal forms of all its
proper divisors have been computed.

A degree-reverse-lexicographical Gr\"obner basis admits to find roots of
the initial polynomial system by the sequential solving of univariate
polynomial equations. Given a univariate polynomial, we tried first to
factorize it and used the built-in factorization routines of computer
algebra system Reduce 3.7~\cite{Reduce} for this purpose. If the
factorization failed to give exact roots we used a special software
package ROOTs written on the top of PARI-GP system~\cite{PARI} to find the
roots numerically for the factors obtained.

\section{The Results for Polynomial Potentials with $q \leq 3$}

\subsection{Sextic QES Oscillator with $q=1$ and Any $N$}

Starting from the first nontrivial sextic-oscillator potential
(\ref{SExta}) with $q=1$ and with the binding energies
re-parametrized in accord with eq. (\ref{recipe}) where $s_1=s$,
 \ben
 E =  \frac{1}{2}\frac{g_1}{\sqrt{g_2}}\,D
 +{(64\,g_2)^{1/4}}\,\sqrt{D} \, s\,,
 \een
full attention must be paid to the selfconsistency problem
represented by the set of equations (\ref{trap}). At every $N$, its
first nontrivial $q=1$ version
 \begin{equation}
 \label{trapegy}
 \left( \begin{array}{cccccc}
 s& 1& &&&\\
 N-1 & s &2 & && \\
  & N-2 & s & 3 &&\\
 &  &\ddots & \ddots & \ddots & \\
& &&2& s & N-1     \\&& && 1 & s
\end{array} \right)
 \left( \begin{array}{c}
 {p}_0\\
 {p}_1\\
\vdots \\
 {p}_{N-2}\\
 {p}_{N-1}
\end{array} \right)
=
0\
 \end{equation}
has the form of an asymmetric eigenvalue problem. In standard manner
it leads to the secular equation (\ref{secular}) expressible as the
following sequence of the polynomial conditions,
 \ben
s^3-4\,s=0, \ \ \ \ \ \ \ N = 3,
 \een
 \ben
s^4-10\,s^2+9=0, \ \ \ \ \ \ \ N = 4,
 \een
 \ben
s^5-20\,s^3+64\,s=0, \ \ \ \ \ \ \ N = 5,
 \een
etc. By mathematical induction, all the infinite hierarchy of these
equations has been recently derived and solved in ref. \cite{RS}.

Quite remarkably, all of the {\em real} (i.e., ``physical") energy
roots $ s=s^{(j)}$ proved to be equal to integers. Moreover, all of
them may be determined by the single and compact formula
 \be
 s=s^{(j)}=-N-1+2j,\ \ \ \ j = 1, 2, \ldots, N.
 \label{sextar}
 \ee
One imagines that all the coefficients
$p_n^{(j)}$ may be normalized to integers,
 \ben
 p_0^{(1)}=1, \ \ \ \ \ \ \ \ \ N = 1,
 \een
 \ben
 p_0^{(1)}=
 p_1^{(1)}=
 p_0^{(2)}=
 - p_1^{(2)}=1,
 \ \ \ \ \ \ \ \ \ N = 2,
 \een
 \ben
 p_0^{(1)}=
 p_2^{(1)}=
 p_0^{(2)}=
 - p_2^{(2)}=
 p_0^{(3)}=
 p_2^{(3)}=1,\ \ \
 p_1^{(1)}=
 - p_1^{(3)}=2,
 \ \ \
 p_1^{(2)}=0, \ \ \  \ N = 3,
 \een
etc.

The first result of our subsequent computations using the symbolic
manipulation techniques proved equally encouraging since we succeeded
in compactification of the set of the above recurrent solutions to
the single leading-order form of the related wave functions,
 \ben
 \psi^{(j)}(r) =r^{\ell+1}\,
 \left(
 1+\frac{r^2}{\mu}
 \right )^{N-j}\,
 \left(
 1-\frac{r^2}{\mu}
 \right )^{j-1}\,
 \exp \left (- \frac{1}{2}{\alpha}_0 r^2 -
  \frac{1}{4}{\alpha}_1 r^4
 \right )\,, \een
 \be\ \ \ \ \ \ \ \ \ \ \ \
 \ \ \ \ \ \ \ \ \
 \ \
 \ \ \ \ \
 \ \ \ \ \ j= 1, 2, \ldots, N\,.
  \label{anesex}
 \ee
A few more comments may be added. Firstly, the large and  degenerate
nodal zeros in eq. (\ref{anesex}) are a mere artifact of the
zero-order construction. This means that the apparently interesting
exact summability of all the separate ${\cal O}(r^2/\mu)$ error terms
is not too relevant, indeed. Although it leads to the zero-order
nodes at $r = {\cal O}(\sqrt{\mu})= {\cal O}(D^{1/4})$, these nodes
have no real physical meaning.

Secondly, the leading-order perturbative approximation provides a
reliable information about the energies. They are asymptotically
degenerate, due to the large overall shift of the energy scale as
explained in section \ref{reasons}. In addition, the next-order
corrections may be easily obtained by the recipes of the textbook
perturbation theory. As long as the coefficients $p_n$ are defined in
integer arithmetics, the latter strategy gives, by construction, all
the above-mentioned energy corrections without any rounding errors in
a way outlined in more detail in ref. \cite{RS}.

In the other words, we may say that formula (\ref{anesex}) may either
be truncated to its leading-order form $ \psi^{(j)}(r) =r^{\ell+1}\,
\exp \left (- \lambda^{(2)}(r) \right )$ or, better, its full form
may be used as a generating function which facilitates the explicit
evaluation of the coefficients $p_n^{(j)}$. In comparison, both the
oversimplified harmonic oscillator and the $q=1$ wave functions may
be characterized by the similar coordinate dependence which becomes
spurious (i.e., dependent on the selected normalization) everywhere
beyond the perturbatively accessible domain of $r$.

The energies specified by eq. (\ref{sextar}) form an amazingly
regular multiplet. A natural question arises whether a similar
regularity could re-emerge at the larger integer indices $q> 1$. We
are now going to demonstrate that in spite of the growth of the
technical obstacles in dealing with the corresponding key equation
(\ref{trap}), the answer is, definitely, affirmative.

\subsection{Decadic Oscillators with $q=2$
and Any $N$}

The decadic anharmonic oscillator exhibits certain solvability
features which motivated its deeper study in non-Hermitian context
\cite{decadic}. The changes of variables make this oscillator very
closely related to the common quartic problem \cite{Dubna,quartic}.
Paying attention to the $D \gg 1$ domain and abbreviating the
parameters $s_1=s$ and $s_2=t$ of the respective decadic-oscillator
energy and coupling in eq. (\ref{recipe}), we arrive at the
four-diagonal version of our solvability condition (\ref{trap}) at
$q=2$,
 \begin{equation}
 \label{trapegyse}
 \left( \begin{array}{ccccccc}
 s& 1& &&&&\\
 t & s &2 & &&& \\
   N-1 & t& s & 3 &&&\\
 &  N-2 & t& s & 4 &&\\
 & & \ddots &\ddots & \ddots & \ddots & \\
& &&3&t& s & N-1
 \\&&& & 2 & t&s
 \\&&& && 1 & t
\end{array} \right)
 \left( \begin{array}{c}
 {p}_0\\
 {p}_1\\
\vdots \\
 {p}_{N-2}\\
 {p}_{N-1}
\end{array} \right)
=
0\ .
 \end{equation}
This is the first really nontrivial equation of the class
(\ref{trap}). In order to understand its algebraic structure in more
detail, let us first choose the trivial case with $N=2$ and imagine
that the resulting problem (with $p_1 \neq 0$ due to the definition
of $N$) may  be solved by the determination of the unknown ratio of
the wave-function coefficients $p_0/p_1=-t$ from the last line, and
by the subsequent elimination of $t= 1/s$ using the first line. The
insertion of these two quantities transforms the remaining middle
line into the cubic algebraic equation $ s^3=1$ with the single real
root $s=1$.

The next equation at $N=3$ is still worth mentioning because it shows
that the strategy accepted in the previous step is not optimal.
Indeed,
the same elimination of $p_1/p_2=-t$ and of $p_0/p_2= (t^2-s)/2$ from
the third line leads to the apparently ugly result
 \ben
 \ba
 st^2-s^2-2t=0,\\
 t^3-3st+4=0 .\ea
 \een
An alternative strategy starting from the
elimination of $p_0$ and $p_2$ leads to the much more symmetric pair
of the conditions
 \ben
 \ba
 t^2-s^2t+2s=0,\\
 s^2-t^2s + 2t=0 \ea
 \een
the respective pre-multiplication of which by $t$ and $s$ gives the
difference $t^3=s^3$. This means that $t=\varepsilon\,s$ where the
three eligible proportionality constants exist such that
$\varepsilon^3=1$. Thus, our problem degenerates to a quadratic
equation with the pair of the real roots $s = t = s^{(1,2)}$ such
that
 \be
 s^{(1)}=2, \ \ \ \ s^{(2)}=-1.
 \label{same}
 \ee
The ``ugliness" of the procedure of elimination is inessential as
long as we can produce the results by any ``brute-force" symbolic
manipulations on the computer \cite{onitri}.

Once we encountered the limitations of the naive algorithms, we were
forced to pay attention to all the above-described sophistications of
our algorithms. Fortunately, this overall strategy proved successful.
Using the methods described in preceding sections we revealed that
the step-by-step elimination of the redundant unknowns gives the best
form of the results when one uses the Janet bases.

One of the main and most important byproducts of our approach is that
the resulting final effective or ``secular" polynomial equations for
the single unknown quantity $s$ depend in practice on its power
$s^{q+1}$ only. The clear illustration is provided by the present
$q=2$ case at $N=3$ giving the rule
 \be
 s^6-7\,s^3-8=0, \ \ \ \ \ N = 3.
 \label{itself}
 \ee
This equation possesses the same complete set of the real roots
(\ref{same}) of course. Still, what is important is that {\em any} root
$r=s^3_0$ of eq. (\ref{itself}) itself still represents just the third
(and in the more general cases, $(q+1)$st) power of the final relevant
quantity with the physical meaning of the energy. Thus, we still have to
solve the relation $r=s^3_0$ where merely the value of $r$ is known and
where, therefore, {\em at most one} final parameter $s_0$ is real (=
acceptable).

One may conclude that the real energies of the ``strongly spiked"
decadic oscillator are very easily determined even without a detailed
specification of an ``optimal" elimination pattern, and that it is
very easy to get rid of the redundant non-real roots $s_0$ at the
very end of the algorithm. Of course, the numerous complex roots
should not be discarded {\it a priori} as they might prove important
in some other applications like a systematic computation of the
corrections \cite{RS} which were not mentioned in our present paper
at all.

Our conclusions extracted at $N=3$ are confirmed at the next $N$
leading to the effective polynomial equation
 \ben
 s^{10} - 27\,s^7 + 27\,s^4 - 729\,s = 0\,, \ \ \ N = 4\,.
 \een
Being tractable by our newly developed computer software and playing
still the role of a test, it results in the set of the mere two real
roots again,
 \ben
 s^{(1)}=3, \ \ \ \ s^{(2)}=0,
   \ \ \ \ \ \ \ N = 4, \ \ \ q = 2.
  \een
One finds that the $q=2$ problem may be reduced to a single
polynomial equation with $\left ( \ba N+1\\2 \ea \right )$ complex
roots $s$ at any $N$. The explicit calculations may be summarized in
a statement that {\em all} the general physical (i.e., real) spectrum
of energies proves to be quite rich and appears described by the
closed and amazingly simple and transparent formula again,
 \be
 s^{(j)}=N+2-3j, \ \ \ \ \ \ j = 1, 2, \ldots,
 j_{max}, \ \ \ \ \ \ \ j_{max}=
 entier \left [ \frac{N+1}{2}
 \right ]\ .
 \label{empir}
 \ee
After one applies our Janet-basis procedure at the higher and higher
dimensions $N$, one repeatedly arrives at the confirmation of the
$N-$independent empirical observation (\ref{empir}) and extends it by
another rule that at all the values of the dimension $N$, there exist
only such real roots that $s^{(j)}=t^{(j)}$. This means that each
''solvability admitting" real energy $s$ requires, purely
constructively, the choice of its own ''solvability admitting" real
coupling constant $t$.

\subsection{Oscillators with $q=3$
and Their Solution at Any $N$}

At $q=3$ we have to solve the five-diagonal eq. (\ref{trap}),
 \begin{equation}
 \label{trapegydr}
 \left( \begin{array}{cccccccc}
 r& 1& &&&&&\\
 s& r& 2& &&&&\\
 t & s &r&3 & &&& \\
   N-1 & t& s & r&4 &&&\\
 &  N-2 & t& s &r&5 &&\\
 & & \ddots &\ddots &\ddots & \ddots & \ddots & \\
 &&&4&t& s &r& N-1
 \\
 &  &&&3&t& s & r
 \\&&&& & 2 & t&s
 \\&&&& && 1 & t
\end{array} \right)
 \left( \begin{array}{c}
 {p}_0\\
 {p}_1\\
\vdots \\
 {p}_{N-2}\\
 {p}_{N-1}
\end{array} \right)
=
0\
 \end{equation}
which may be reduced, by means of the similar symbolic computations
as above, to the single polynomial problem
 \ben
 t^{9}-12\,t^{5}-64\,t=0
 \een
at $N=3$, to the next similar condition
 \ben
 t^{16}-68\,t^{12}-442\,t^8-50116\,t^4+50625=0
 \een
at $N=4$, to the conditions of vanishing of the secular polynomial
 \ben
 t^{25}-260\,t^{21}+7280\,t^{17}-1039040\,t^{13}-152089600\,t^9
 +2030239744\,t^5+10485760000\,t
 \een
at $N=5$, or to the perceivably longer equation
 \ben
t^{36}-777\,t^{32}+135716\,t^{28}-17189460\,t^{24}-3513570690\,t^{20}
 -\een \ben-1198527160446\,t^{16}
+103857100871252\,t^{12}+873415814269404\,t^8+\een
\ben+74500845455535625\,t^4 -75476916312890625=0
 \een
at $N=6$ etc. These computations represent a difficult technical task
but at the end they reveal again a clear pattern in the structure of
the secular polynomials as well as in their solutions. One arrives at
the similar final closed formulae as above. Now one only deals with
more variables so that we need two indices to prescribe the complete
classification scheme
 \ben
 s=s^{(j)}= N+3-4j, \ \ \ \ \ \ \ \
 \een
 \be
 r =
 r^{(j,k)}=t =
 t^{(j,k)}= -N-3+2j+2k,
 \ee
 \ben
  \ \ \ \ \ \ k = 1, 2, \ldots,
 k_{max}(j), \ \ \ \ \ \ k_{max}(j)=
 N+2-2j\ ,
 \een
 \ben
 \ \ \ \ \ \ \ \ \ \
 j = 1, 2, \ldots,
 j_{max}\ ,
  \ \ \
  \ \ \ \ j_{max}=
 entier \left [ \frac{N+1}{2}
 \right ]\ .
 \een
We may re-emphasize that all the real roots share the symmetry $r=t$
but admit now a different second root $s$. The physical meaning of
these roots is obvious. Thus, the energies of the oscillations in the
polynomial well
 \ben
 V^{(q=3,k=1)}(r) =  a\,r^2+b\,r^4+\ldots + g\,r^{14}
 \een
will be proportional to the roots $ r^{(j,k)}$. After the change of
variables, the roots $s^{(j)}$ will represent energies for the
alternative, ``charged" polynomial potentials
 \ben
 V^{(q=3,k=2)}(r) = \frac{e}{r}+ a\,r+b\,r^2+\ldots + f\,r^6
 \een
etc \cite{classif}.

\section{The Results with $q = 4$ and $q= 5$ for $N \leq N_{max}$}

\subsection{Non-Integer Roots Emerging at $q=4$ and $N \leq 6$}

In our present formulation of the problem (\ref{trap}), we denote the
descending diagonals as $s_{m}$ with $m=1,2,3,4$ and get the equation
 \begin{equation}
 \label{tregydr}
 \left( \begin{array}{cccc}
 s_1& 1& &\\
 s_2& \ddots&\ddots& \\
  s_3&\ddots&\ddots&N-1\\
 s_4&\ddots&\ddots& s_1\\
 N-1&\ddots&\ddots&s_2\\
 &\ddots&\ddots&s_3\\
 &&1&s_4
\end{array} \right)
 \left( \begin{array}{c}
 {p}_0\\
 {p}_1\\
\vdots \\
 {p}_{N-1}
\end{array} \right)
=
0\ .
 \end{equation}
Its systematic solution does not parallel completely the
above-described procedures. In fact, the reduction of the problem to
the search for the roots of a single polynomial secular equation
$P(x)=0$ (in the selected auxiliary variable $x = -s_4$) enables us
only to factorize $P(x)$ on an extension of the domain of integers,
  \ben
  P(x)= \left (x+3\right )
 \left (2\,x+1-\sqrt
{5}\right ) \left (2\,x+1+\sqrt {5}\right )
 \een
 \ben
  \left (2\,{x}^{2}-3\,x+3\,\sqrt
{5}x+18\right ) \left (2\,{x}^{2}-3\,x-3\,\sqrt {5}x+18\right )
 \een
 \ben
 \left
(2\,{x}^{2}-3\,x- \sqrt {5}x+8+2\,\sqrt {5}\right )\left
(2\,{x}^{2}-3\,x+\sqrt {5}x+8-2 \,\sqrt {5}\right )
 \een
 \ben
 \left
({x}^{2}+x+\sqrt {5}x+4+\sqrt {5}\right )
  \left ({x}^{2}+x-\sqrt
{5}x+4-\sqrt {5}\right )
 \een
 \ben
 \left (-2\,\sqrt {5}+8- 3\,x+3\,\sqrt
{5}x+2\,{x}^{2}\right )
 \left (2\,\sqrt {5}+8-3\,x-3\, \sqrt
{5}x+2\,{x}^{2}\right )
 \een
 \ben
 \left (-2\,\sqrt {5}+8+7\,x-\sqrt {5}x+2
\,{x}^{2}\right )
 \left (2\,\sqrt {5}+8+7\,x+\sqrt {5}x+2\,{x}^{2}
\right )
 \een
 \ben
 \left (2\,{x}^{2}+2\,x+3-\sqrt {5}\right )
 \left (2\,{x}^{2}+2
\,x+3+\sqrt {5}\right )
 \een
 \ben
 \left (\sqrt {5}+3-3\,x-\sqrt {5}x+2\,{x}^{2}
\right )\left (-\sqrt {5}+3-3\,x+\sqrt {5}x+2\,{x}^{2}\right )
 \een
 \ben
 \left
(2 \,\sqrt {5}+8-3\,x+\sqrt {5}x+2\,{x}^{2}\right )
 \left (-2\,\sqrt
{5}+8 -3\,x-\sqrt {5}x+2\,{x}^{2}\right )\ .
 \een
From this lengthy formula it follows that we get
 \ben
 s_4^{(1)}=3, \ \ \ \
 s_4^{(2)}=\frac{\sqrt{5}+1}{2} \approx 1.618, \ \ \ \
 s_4^{(3)}=\frac{\sqrt{5}-1}{2} \approx -0.618\ .
 \een
There only exist these three real roots $s_4$ in this case.

The similar computerized procedure gave us the real roots also at
$N=5$ and $N=6$. The details may be found in ref. \cite{papI}. The
inspection of these results leads to the conclusion that $s_2=s_3$
and $s_1=s_4$. We did not succeed in an application of our algorithms
beyond $N=6$ yet. The reason is that even the $N=5$ version of eq.
(\ref{tregydr}) in its reduction to the condition
 \ben
 x^{70}-936\,x^{65}+67116\,x^{60}-95924361\,x^{55}-74979131949\,x^{50}
 +8568894879002\,x^{45}-
 \een
 \ben
 \ldots -17459472274501870222336\,x^5+142630535951654322176=0
 \een
of the vanishing auxiliary polynomial required a fairly long
computation for its (still closed and compact) symbolic-manipulation
factorization summarized in Table~1 of ref. \cite{papI}.

\subsection{A Mind-Boggling Return of Integer Roots at $q=5$}

\subsubsection{$N=6$}

At $q=5$ and $N=6$ the symbolic manipulations using the Gr\"{o}bner
bases \cite{Buch85} generate the secular polynomial in $x=s_5$
which has the slightly deterring form
 \ben
x^{91}-16120\,x^{85}+49490694\,x^{79}-286066906320\,
x^{73}-3553475147614293\,x^{67}-
 \een
 \ben
 \ldots
 -319213100611990814833843025405983064064000000\,x=0\ .
  \een
Fortunately, it proves proportional to the polynomial with the mere
equidistant and simple real zeros,
 \ben
 P_1^{(6)}(x)=
x\left (x^2-1\right )\left (x^2-2^2\right )\left (x^2-3^2\right
)\left (x^2-4^2 \right )\left (x^2-5^2\right )\ .
 \een
The rest of the secular polynomial is a product of the other two
elementary and positive definite polynomial factors
 \ben
 P_2^{(6)}(x)=\prod_{k=1}^{2}\,
 \left (
 x^2-3k\,x+3k^2
 \right )
 \left (
 x^2+3k^2
 \right )
 \left (
 x^2+3k\,x+3k^2
 \right )
 \een
and
 \ben
 P_{3}^{(6)}=
 \prod_{k=1}^{5}\,
 \left (
 x^2- k\,x+k^2
 \right )
 \left (
 x^2+ k\,x+k^2
 \right ),
 \een
with another positive definite polynomial
 \ben
 P_{4}^{(6)}=
 \prod_{k=1}^{12}\,
 \left (
 x^2- b_k\,x+c_k
 \right )
 \left (
 x^2+ b_k\,x+c_k
 \right )\
 \een
where the structure of the two series of coefficients (see their list
in ref. \cite{papI}) is entirely enigmatic.

The subsequent symbolic manipulations reveal a symmetry $s_2=s_4$ and
$s_1=s_5$ of all the real eigenvalues. In the $N=6$ pattern
summarized in ref. \cite{papI} we recognize a clear indication of a
tendency of a return to the transparency of the $q \leq 3$ results
which may be written and manipulated in integer arithmetics. For
obtaining a deeper insight we must move to the higher $N$.

\subsubsection{$N=7$}

One should note that in spite of its utterly transparent form, the
latter result required a fairly long computing time for its
derivation. One encounters new technical challenges here.
Indeed, the comparison of the $N=6$ secular polynomial equation with
its immediate $N=7$ descendant
 \ben
x^{127}-60071\,x^{121}+1021190617\,x^{115}-11387407144495\,x^{109}-\ldots+
c\,x\cdot
 10^6 =0
 \een
shows that the last coefficient
 \ben
c= 125371220122726667620073789326658415654595883041274311330630729728
 \een
fills now almost the whole line. This case failed to be tractable by
our current computer code and offers the best illustration of the
quick growth of the complexity of the $q\geq 5$ constructions with
the growth of the QES dimension parameter $N$.

Fortunately, we are still able to keep the trace of the pattern
revealed at $N=6$. Indeed, our new secular $N=7$ polynomial
factorizes again in the product of the four factors $P_j(x)$, $j = 1,
2, 3, 4$ where only the first one has the real zeros,
 \ben
 P_1^{(7)}(x)=P_1^{(6)}(x)\cdot
\left (x^2-6^2\right )\ .
 \een
The further three factors fit the structure of their respective
predecessors very well,
 \ben
 P_2^{(7)}(x)=P_2^{(6)}(x)\cdot
 \left (
 x^2-9\,x+27
 \right )
 \left (
 x^2+27
 \right )
 \left (
 x^2+9\,x+27
 \right )
 \een
and
 \ben
 P_{3}^{(7)}=P_{3}^{(6)}\cdot
  \left (
 x^2- 6\,x+36
 \right )
 \left (
 x^2+ 6\,x+36
 \right )
 \een
while
 \ben
 P_{4}^{(7)}=P_{4}^{(6)}\cdot
 \prod_{k=1}^{6}\,
 \left (
 x^2- f_k\,x+g_k
 \right )
 \left (
 x^2+ f_k\,x+g_k
 \right )\ .
 \een
The subscript-dependence of the new coefficients may be found in ref.
\cite{papI} again. The key importance of the explicit knowledge of
these coefficients lies in the possibility of a rigorous proof that
the related roots are all complex and, hence, irrelevant from our
present point of view.

\subsubsection{$N=8$ and $N=9$}

The growth of the degree of our secular univariate polynomials makes it
quite difficult to move too far with $N$. One may be more explicit in this
respect: In place of the numerous irregularities encountered at $q=4$, we
may now be surprised by the re-emergence of the following {\em closed} and
very transparent elementary formula for the $q=5$ ``energies",
 \be
 s_5 \in (-N+1, -N+2, \ldots, N-2, N-1)\,.
 \label{extension}
 \ee
which is valid again for {\em any} integer $N$ in a way which parallels
and complements the above-mentioned results which were available and
published in our previous papers \cite{RS}, \cite{quartic} and \cite{papI}
for the Magyari's $D \gg 1$ potentials with $q=1$, $q=2$ and $q=3$,
respectively. In this context, their extension (\ref{extension}) is a
brand new result which has not been published yet. Its unexplained
equidistance property may be added to the list of the unresolved puzzles
related to the Magyari equations. Indeed, the equidistance exemplified by
eq. (\ref{extension}) would reflect a hidden algebra in linear cases but
what is most intriguing here is the fact that the present Magyari
equations are {\em non-linear}!

Another challenging feature of the problem lies in its exact
solvability which is based on the factorization of polynomials of a
very large degree ${\cal D}$ which grows, moreover, very quickly with
$N$. Empirically, this degree may be even specified by the closed
formula at $q=5$ where ${\cal D} = 3N^2-3N+1$ in a way illustrated by
the next two explicit secular equations
 \be
  s_{5}^{169}-186238\,s_{5}^{163}+11768813199\,s_{5}^{157}-\ldots =
  0,
  \ \ \ \ N=8,
 \ee
 \be
 s_{5}^{217}-502386\,s_{5}^{211}+94933635261\,s_{5}^{205}-\ldots = 0,
  \ \ \ \ \ \
 N=9.
 \ee
On the basis of these observations we may conclude that an overall
pattern of the smooth $N-$dependence of the equations survives, {\it
mutatis mutandis}, smoothly the transition to the higher $N$. One can
also prove (at least up to $N \leq 9$ at present) by construction
that {\em all} the other factors of the secular polynomial have an
elementary quadratic-polynomial form and remain positive for all the
real ``re-scaled energies" $s_{5}$. Their coefficients are elementary
(we skip the examples here) so that the strict proof that they
possess no real zeros is also elementary and very quick (one just
evaluates the discriminants). A full parallelism between all $N\leq
9$ is achieved and we might conjecture, on this background, the
possibility of its extension to all the non-negative integers $N$. A
strict proof of this conjecture could probably be based on
mathematical induction but we do not feel it really urgent at the
moment.

\section{Summary}

It is rather amusing to imagine that the {\em majority} of
quantitative predictions in nuclear, atomic, molecular and condensed
matter physics must rely on a more or less purely numerical model.
The completely non-numerically tractable quantum systems are rare
though, at the same time, useful and transparent (cf., e.g., the
above-mentioned description of vibrations in molecules mimicked by
harmonic oscillators). In our present paper we revealed that in the
domain of the large spatial dimension $D \gg 1$, the class of the
exactly solvable models becomes, in a certain sense, broader. Thus,
one might call {\em all} the polynomially anharmonic oscillators
``asymptotically solvable".

This is an important and also not yet fully appreciated observation
obtained due to the lasting advancement of the computer algebra and
related software as described in more detail in Sections 4 and 5. A
fairly universal apparatus of these sections was reported in close
connection with its application to our Magyari-type equations
(\ref{trap}).

In a certain perspective we found new closed solutions of
Schr\"{o}dinger equation with polynomial potentials in the domain of
the large angular momenta $\ell \gg 1$ where alternative techniques
are also available (cf., e.g., their review \cite{Bjerrum} and/or
very recent discussion \cite{dis}). Our results revealed the
existence and provided the construction of certain {fairly large}
multiplets of ``exceptional" $\ell \gg 1$ bound states for a very
broad class of polynomial oscillators. We believe that they might
find an immediate application in some phenomenological $D \gg 1$
models.

From the mathematical point of view, the most innovative and
characteristic feature of our new $D \gg 1$ QES multiplets lies in the
existence of the new {\em closed and compact} formulae for the QES
energies and/or couplings {\em at all $N$}. For this reason, the
corresponding partially solvable polynomial oscillator Hamiltonians
$H_0^{(q,N)}$ might even be understood as lying in the QES class as its
new and fairly specific subclass.

Due to an exceptional transparency of our constructions of
$H_0^{(q,N)}$, a facilitated return to the ``more realistic" finite
spatial dimensions $D={\cal O}(1)$ might prove tractable by
perturbation techniques. Two reasons may be given in favor of such a
strategy. First, due to the specific character of our present
``unperturbed" spectra {\em and} eigenvectors, the perturbation
algorithm might be implemented {\em in integer arithmetics} (i.e.,
without rounding errors) in a way outlined, preliminarily, in ref.
\cite{RS} at $q=1$. Second, the evaluation of the few lowest orders
might suffice. This expectation follows from the enhanced flexibility
of the available zero-order Hamiltonians. {\it A priori}, a better
convergence of the corrections might be expected to result from a
better quality of a ``guaranteed smallness" of the difference between
a given Hamiltonian $H$ at a finite $D$ and one of its present
$D=\infty$ QES approximants $H_0$.

\section{Acknowledgements}
The contribution of V.G. and D.Y. was partially supported by the
grant 01-01-00708 from the Russian Foundation for Basic Research and
grant 2339.2003.2 from the Russian Ministry of Industry, Science and
Technologies. M.Z. appreciates the support by the grant Nr. A
1048302 of GA AS CR.

\end{document}